\documentclass{revtex4}
\usepackage{amssymb}
\usepackage{amsfonts}
\usepackage{amsmath}

\input{tcilatex}

\begin{document}

\title{Necessary conditions for having wormholes in f(R) gravity}
\author{S. Habib Mazharimousavi}
\email{habib.mazhari@emu.edu.tr}
\author{M. Halilsoy}
\email{mustafa.halilsoy@emu.edu.tr }
\affiliation{Department of Physics, Eastern Mediterranean University, G. Magusa, north
Cyprus, Mersin 10, Turkey. }

\begin{abstract}
For a generic $f(R)$ which admits a polynomial expansion of at least third
order (i.e. $\frac{d^{3}f}{dR^{3}}\neq 0$) we find the near-throat wormhole
solution. Necessary conditions for the existence of wormholes in such $f(R)$
theories are derived for both zero and non-zero matter sources. A particular
choice of energy-momentum reveals that the wormhole geometry satisfies the
weak energy condition (WEC). For a range of parameters even the strong
energy condition (SEC) is shown to be satisfied.
\end{abstract}

\maketitle

\section{Introduction}

Black hole solutions in modified theories of gravity (namely $f(R)$ gravity)
has received attentions in recent years \cite{1}. Parallel to black holes
there are studies on wormhole solutions in $f(R)$ gravity as well \cite{2}.
Upon derivation of the necessary conditions for the existence of black holes
in $f(R)$ gravity \cite{3}, it will be in order to explore similar
conditions for wormholes in the same theory. This amounts to make series
expansions for functions around the throat radius ($r=r_{0}$) of the
wormhole. That is, any function $h(r)$ can be expressed as $%
h(r)=h(r_{0})+h^{\prime }(r_{0})(r-r_{0})+\mathcal{O}((r-r_{0})^{2}).$ Such
expansions are carried out for the Ricci scalar $R,$ $f(R),$ $F=\frac{df}{dR}
$ and higher derivatives, as well as the pressure and density of the
energy-momentum tensor. Let us remind that even in the absence of external
sources the curvature sources of $f(R)$ theory remain intact. The reason for
resorting these expansions is entirely physical: near the throat radius of a
wormhole designated by $r-r_{0}=x>0,$ ($x^{2}\ll 1$) an observer doesn't
feel anything unusual so that the functions specifying all salient features
are expressible in analytic functions. Once these expansions are substituted
into Einstein's equations and zeroth, first and higher order terms in $%
\left\vert r-r_{0}\right\vert $ are identified automatic constraints emerge.
These will comprise the necessary conditions for the existence of wormhole
throats in any viable $f(R)$ theory as a result of such a "\textit{%
near-throat test}". At the zeroth order the conditions are easily tractable
and yield tangible results to say enough about an $f(R)$ whether it admits a
throat or not. We must note that the higher order terms, even the first one
turns out to yield rather complicated relations in implementing the test. In
this study we concentrate on an $f(R)$ theory that admits at least a
non-zero third order derivative, i.e. $\frac{d^{3}f}{dR^{3}}\neq 0,$ while
all the rest will be dictated by Einstein's equations. In this study our
ansatz metric involves two unknown metric functions to be determined: the
redshift function $\Phi \left( r\right) $ and the shape function $b\left(
r\right) $. Upon expansions aforementioned in powers of ($r-r_{0}$) there
will be severe restrictions on these functions. The simpler case consists of
choosing $\Phi \left( r\right) $ as constants, however, herein this will not
be our strategy. One thing observed is that in the absence of matter sources
the weak energy condition (WEC) is violated in the construction of
wormholes. To overcome this problem we introduce external energy-momentum
and search for the satisfaction of the energy conditions. We find that WEC
holds true while the strong energy condition (SEC) becomes valid under more
stringent regulations. Integration of $f(R)$ incorporates constants that are
related to the cosmological constant and with / without that constant
validity of our energy conditions remains still valid. After we determine
the redshift and shape functions analytically we embed our geometry into the
($r,z,\phi $) sector with the embedding function $z=\pm r_{0}\cosh
^{-1}\left( \frac{r}{r_{0}}\right) .$ The shape of our traversable wormhole
is depicted in Fig. 1 for the particular throat radius $r_{0}=1.$

The paper is organized as follows. In Sec. II we introduce our formalism of
expansion around the throat in a generic $f(R)$ theory. Section III extends
/ generalizes the formalism in the presence of external matter sources. With
Conclusion in Section IV we complete the paper.

\section{The Formalism}

We start with a general action for $f(R)$ gravity written as%
\begin{equation}
S=\frac{1}{2\kappa }\int \sqrt{-g}f\left( R\right) d^{4}x
\end{equation}%
in which $\kappa =8\pi G=1,$ and $f\left( R\right) $ is a real arbitrary
function of the Ricci scalar $R.$ Obviously, in the presence of physical
sources the action is to be supplemented by a matter part $S_{M}.$ The $4-$%
dimensional standard form of the wormhole line element in spherical symmetry
is given by \cite{4}%
\begin{equation}
ds^{2}=-e^{2\Phi \left( r\right) }dt^{2}+\frac{1}{\left( 1-\frac{b\left(
r\right) }{r}\right) }dr^{2}+r^{2}\left( d\theta ^{2}+\sin ^{2}\theta
d\varphi ^{2}\right)
\end{equation}%
where $\Phi \left( r\right) $ and $b\left( r\right) $ are the redshift and
shape functions, respectively. The throat of the wormhole is located at $%
r=r_{0},$ at which $b\left( r_{0}\right) =r_{0}$ and radial coordinate $r$
is larger than $r_{0}.$ We note that $\Phi \left( r\right) $ and $b\left(
r\right) ,$ in addition to $R,$ $f(R)$ and its higher derivatives should
also satisfy constraints to have a traversable wormhole. These conditions
for $b(r)$ and $\Phi \left( r\right) $ are: i) $b^{\prime }\left(
r_{0}\right) <1,$ ii) $b\left( r\right) <r$ for $r>r_{0}$ and iii) $e^{2\Phi
\left( r\right) }$ must not have any root (horizon) i.e., $\Phi \left(
r\right) $ must be finite everywhere.

Variation of the action with respect to the metric yields the field equation%
\begin{equation}
FR_{\mu }^{\nu }-\frac{f}{2}\delta _{\mu }^{\nu }-\nabla ^{\nu }\nabla _{\mu
}F+\delta _{\mu }^{\nu }\square F=0
\end{equation}%
in which $\square =\nabla ^{\mu }\nabla _{\mu }=\frac{1}{\sqrt{-g}}\partial
_{\mu }\left( \sqrt{-g}\partial ^{\mu }\right) $ and $\nabla ^{\nu }\nabla
_{\mu }h=g^{\lambda \nu }\nabla _{\lambda }h_{,\mu }=g^{\lambda \nu }\left(
\partial _{\lambda }h_{,\mu }-\Gamma _{\lambda \mu }^{\beta }h_{,\beta
}\right) $. The field equations explicitly read as%
\begin{equation}
FR_{t}^{t}-\frac{f}{2}+\square F=\nabla ^{t}\nabla _{t}F
\end{equation}%
\begin{equation}
FR_{r}^{r}-\frac{f}{2}+\square F=\nabla ^{r}\nabla _{r}F
\end{equation}%
\begin{equation}
FR_{\theta }^{\theta }-\frac{f}{2}+\square F=\nabla ^{\theta }\nabla
_{\theta }F
\end{equation}%
which are independent. Note that the $\varphi \varphi $ equation is
identical with $\theta \theta $ equation. By adding the four equations
(i.e., $tt$, $rr$, $\theta \theta $ and $\varphi \varphi $) we find 
\begin{equation}
FR-2f+3\square F=0
\end{equation}%
which is the trace of Eq. (3).

Our method to solve these equations is as follows: First, we consider the
Ricci scalar as a series about the throat i.e., 
\begin{equation}
R=R_{0}+R_{0}^{\prime }x+\frac{1}{2}R_{0}^{\prime \prime }x^{2}+....
\end{equation}%
in which $x=r-r_{0}>0$. Herein and in the sequel a prime stands for the
derivative with respect to $r$ and a sub $0$ implies that such a quantity is
evaluated at the throat. Next, we expand all other functions involved in the
field equations i.e., 
\begin{eqnarray}
\Phi &=&\Phi _{0}+\Phi _{0}^{\prime }x+\frac{1}{2}\Phi _{0}^{\prime \prime
}x^{2}+... \\
b &=&b_{0}+b_{0}^{\prime }x+\frac{1}{2}b_{0}^{\prime \prime }x^{2}+... \\
f &=&f_{0}+f_{0}^{\prime }x+\frac{1}{2}f_{0}^{\prime \prime }x^{2}+... \\
F &=&\frac{df}{dR}=F_{0}+F_{0}^{\prime }x+\frac{1}{2}F_{0}^{\prime \prime
}x^{2}+... \\
E &=&\frac{d^{2}f}{dR^{2}}=E_{0}+E_{0}^{\prime }x+\frac{1}{2}E_{0}^{\prime
\prime }x^{2}+... \\
H &=&\frac{d^{3}f}{dR^{3}}=H_{0}+H_{0}^{\prime }x+\frac{1}{2}H_{0}^{\prime
\prime }x^{2}+...
\end{eqnarray}%
and finally by equating different powers of $x$ we find all coefficients in
terms of $r_{0},$ $R_{0},$ $R_{0}^{\prime },...$. In zeroth order we find 
\begin{eqnarray}
b_{0} &=&r_{0};\text{ }b_{0}^{\prime }=\frac{1}{3}\left(
r_{0}^{2}R_{0}-1\right) ;\text{ }\Phi _{0}^{\prime }=-\frac{r_{0}^{2}R_{0}+2%
}{r_{0}\left( 4-r_{0}^{2}R_{0}\right) };F_{0}=\frac{r_{0}^{2}f_{0}}{%
r_{0}^{2}R_{0}-2};\text{ \ } \\
E_{0} &=&-\frac{2r_{0}f_{0}}{\left( r_{0}^{2}R_{0}-2\right) R_{0}^{\prime }};%
\text{ \ }f_{0}^{\prime }=\frac{R_{0}^{\prime }r_{0}^{2}f_{0}}{%
r_{0}^{2}R_{0}-2};F_{0}^{\prime }=-2\frac{r_{0}f_{0}}{r_{0}^{2}R_{0}-2};
\end{eqnarray}%
while to first order in $x$ we obtain%
\begin{equation}
\text{ \ }b_{0}^{\prime \prime }=\frac{1}{3r_{0}}\left( R_{0}^{\prime
}r_{0}^{3}+r_{0}^{2}R_{0}-8\right) ;\Phi _{0}^{\prime \prime }=-\frac{%
r_{0}^{4}R_{0}^{2}+8-2r_{0}^{2}R_{0}+2r_{0}^{3}R_{0}^{\prime }}{%
r_{0}^{2}\left( r_{0}^{2}R_{0}-4\right) ^{2}};
\end{equation}%
\begin{equation}
\text{ }H_{0}=2\frac{f_{0}\left( R_{0}^{\prime \prime }r_{0}+R_{0}^{\prime
}\right) }{R_{0}^{\prime 3}\left( r_{0}^{2}R_{0}-2\right) };\text{\ }%
E_{0}^{\prime }=2\frac{f_{0}\left( R_{0}^{\prime \prime }r_{0}+R_{0}^{\prime
}\right) }{R_{0}^{\prime 2}\left( r_{0}^{2}R_{0}-2\right) };\text{ }%
f_{0}^{\prime \prime }=\frac{r_{0}f_{0}\left( R_{0}^{\prime \prime
}r_{0}-2R_{0}^{\prime }\right) }{r_{0}^{2}R_{0}-2}.
\end{equation}%
We note that in our solution $\Phi _{0}$ and $f_{0}$ are not specified. As
one can see from the line element $\Phi _{0}$ can be absorbed into the
redefinition of time $t,$ so we set it to be zero. On the other hand $f_{0}$
is a principal constant which we leave free so that other constants can be
expressed in terms of it i.e., $\frac{F_{0}}{f_{0}}=\frac{r_{0}^{2}}{%
r_{0}^{2}R_{0}-2}$ and so on. Clearly this is just the zeroth order
approximation and it can not be considered for the generic solution.

In order to find the conditions $b\left( r\right) $ must satisfy we start
with $b^{\prime }\left( r_{0}\right) <1,$ which implies%
\begin{equation}
r_{0}^{2}R_{0}<4.
\end{equation}%
The other condition i.e., $b\left( r\right) <r$ for $r>r_{0}$ is
automatically satisfied since%
\begin{equation}
\frac{1}{3}\left( r_{0}^{2}R_{0}-1\right) +\frac{1}{2}b_{0}^{\prime \prime
}x+...<1
\end{equation}%
and for small $x$ it leads to $r_{0}^{2}R_{0}<4$ which is assumed valid. The
last condition which constrains $\Phi \left( r\right) $ to be finite takes
the form 
\begin{equation}
\Phi \left( r\right) =-\frac{r_{0}^{2}R_{0}+2}{r_{0}\left(
4-r_{0}^{2}R_{0}\right) }x-\frac{%
r_{0}^{4}R_{0}^{2}+8-2r_{0}^{2}R_{0}+2r_{0}^{3}R_{0}^{\prime }}{%
2r_{0}^{2}\left( r_{0}^{2}R_{0}-4\right) ^{2}}x^{2}+...
\end{equation}%
This is guaranteed if $x$ is small enough to keep $\frac{x}{\left(
4-r_{0}^{2}R_{0}\right) }$ finite together with $r_{0}^{2}R_{0}<4$.

Now let's consider the energy conditions. If we directly write the field
equations in the form of standard Einstein equation%
\begin{equation}
G_{\mu }^{\nu }=\frac{1}{F}T_{\mu }^{\nu }+\check{T}_{\mu }^{\nu }
\end{equation}%
in which $G_{\mu }^{\nu }$ stands for the Einstein's tensor, with~%
\begin{equation}
\check{T}_{\mu }^{\nu }=\frac{1}{F}\left[ \nabla ^{\nu }\nabla _{\mu
}F-\left( \square F-\frac{1}{2}f+\frac{1}{2}RF\right) \delta _{\mu }^{\nu }%
\right]
\end{equation}%
and $T_{\mu }^{\nu }$ is the external stress energy tensor which in the
present case is zero. Next, we consider the energy density $\rho $ and the
pressure components produced by the geometry of $f(R)$ gravity in the form 
\cite{5}%
\begin{equation}
\check{T}_{\mu }^{\nu }=diag\left[ -\rho \left( r\right) ,p_{r}\left(
r\right) ,p_{t}\left( r\right) ,p_{t}\left( r\right) \right] .
\end{equation}%
After some manipulation we find at the throat (i.e. from the zeroth order
Einstein's equations) 
\begin{equation}
\rho =-\check{T}_{0}^{0}=\frac{r_{0}^{2}R_{0}-1}{3r_{0}^{2}},
\end{equation}%
\begin{equation}
p_{r}=\check{T}_{1}^{1}=-\frac{1}{r_{0}^{2}}
\end{equation}%
and 
\begin{equation}
p_{\theta }=p_{\varphi }=\check{T}_{2}^{2}=\frac{1-r_{0}^{2}R_{0}}{3r_{0}^{2}%
}.
\end{equation}%
It is observed that imposing $1<r_{0}^{2}R_{0}<4,$ makes $\rho $ positive
but $p_{\theta }=p_{\varphi }$ remain negative and in any case $p_{r}$ is
negative. Note that it may be this negative pressure that protects the
wormhole against collapse. From the summation of (25) and (26) it can be
seen easily that the weak energy condition (WEC) which says that $\rho
+p_{i}\geqslant 0$ and $\rho \geqslant 0,$ is violated. In the next section
we may avert this situation by adding external sources.

\subsection{Wormhole supported by a matter source}

In this section we consider a matter source which provides an
energy-momentum tensor of the form%
\begin{equation}
T_{\mu }^{\nu }=diag\left[ -\rho ,p,q,q\right]
\end{equation}%
in which $\rho ,$ $p$ and $q$ are arbitrary functions of $r.$ The field
equations, read now%
\begin{equation}
G_{\mu }^{\nu }=\frac{1}{F}T_{\mu }^{\nu }+\check{T}_{\mu }^{\nu }.
\end{equation}%
The same method as we used in the previous section, i.e., expansion of
quantities about the throat, including $\rho ,$ $p$ and $q$%
\begin{eqnarray}
\rho &=&\rho _{0}+\rho _{0}^{\prime }x+... \\
p &=&p_{0}+p_{0}^{\prime }x+... \\
q &=&q_{0}+q_{0}^{\prime }x+...
\end{eqnarray}%
would lead to the following results in the zeroth order:%
\begin{equation}
\rho _{0}=\frac{1}{2}f_{0}+\frac{F_{0}^{\prime }\left( b_{0}^{\prime
}-1\right) }{2r_{0}}-\frac{\left( b_{0}^{\prime }-1\right) \Phi _{0}^{\prime
}F_{0}}{2r_{0}},
\end{equation}%
\begin{equation}
p_{0}=-\frac{1}{2}f_{0}+\frac{F_{0}\left( b_{0}^{\prime }-1\right) \left(
\Phi _{0}^{\prime }r_{0}+2\right) }{2r_{0}^{2}}
\end{equation}%
and 
\begin{equation}
q_{0}=-\frac{1}{2}f_{0}+\frac{F_{0}\left( b_{0}^{\prime }+1\right) }{%
2r_{0}^{2}}-\frac{F_{0}^{\prime }\left( b_{0}^{\prime }-1\right) }{2r_{0}}.
\end{equation}%
Next, we try to apply WEC which implies $\rho _{0}\geq 0$ and $\rho
_{0}+p_{i0}\geq 0.$ One finds the following conditions which should be
satisfied simultaneously:%
\begin{equation}
\rho _{0}\geq 0\rightarrow f_{0}-\frac{F_{0}^{\prime }\left( 1-b_{0}^{\prime
}\right) }{r_{0}}+\frac{\left( 1-b_{0}^{\prime }\right) \Phi _{0}^{\prime
}F_{0}}{r_{0}}\geq 0,
\end{equation}%
\begin{equation}
\rho _{0}+p_{0}\geq 0\rightarrow r_{0}F_{0}^{\prime }+2F_{0}\leq 0.
\end{equation}%
and 
\begin{equation}
\rho _{0}+q_{0}\geq 0\rightarrow F_{0}\left[ 1+r_{0}\Phi _{0}^{\prime }\frac{%
1-b_{0}^{\prime }}{1+b_{0}^{\prime }}\right] \geq 0
\end{equation}%
in which we used the choice that $b_{0}^{\prime }-1<0$. Next, one may add to
the foregoing conditions $\rho _{0}+p_{0}+2q_{0}\geq 0$ which implies the
strong energy conditions (SEC) and gives%
\begin{equation}
\frac{2b_{0}^{\prime }F_{0}}{r_{0}^{2}}+\frac{F_{0}^{\prime }\left(
1-b_{0}^{\prime }\right) }{2r_{0}}-f_{0}\geq 0.
\end{equation}%
In the case of $\Phi ^{\prime }=0$ \cite{5} one obtains 
\begin{equation}
WEC\rightarrow f_{0}\geq \frac{F_{0}^{\prime }\left( 1-b_{0}^{\prime
}\right) }{r_{0}};\text{ }\frac{F_{0}^{\prime }}{F_{0}}\leq -\frac{2}{r_{0}};%
\text{ }F_{0}\geq 0\text{ (and consequently }F_{0}^{\prime }\leq 0\text{) };
\end{equation}%
and 
\begin{equation}
SEC\rightarrow WEC\text{ and }0\leq \frac{f_{0}}{F_{0}}-\frac{F_{0}^{\prime
}\left( 1-b_{0}^{\prime }\right) }{F_{0}r_{0}}\leq \frac{2b_{0}^{\prime }}{%
r_{0}^{2}}-\frac{F_{0}^{\prime }\left( 1-b_{0}^{\prime }\right) }{2F_{0}r_{0}%
}.
\end{equation}%
We note that although these conditions are very complicated in principle one
can choose the proper parameters to satisfy them. In the next section we
give an exact solution to the problem and with a particular example we shall
justify our prediction.

\section{Generalization}

In this section we study the field equations not at the throat but at any $%
r\geqslant r_{0}.$ This is also a generalization of Ref. \cite{5} in which $%
\Phi =$constant. The field equations are given in Eq.s (4)-(7) and the line
element is considered as (2) while the energy momentum is given by (28). A
detailed calculation would imply the following relations for the components
of energy momentum:%
\begin{equation}
\rho =\frac{f}{2}+\frac{2r\left( \Phi ^{\prime \prime }+\Phi ^{\prime
2}\right) \left( r-b\right) +\left[ -3b+\left( 4-b^{\prime }\right) r\right]
\Phi ^{\prime }}{2r^{2}}F+\frac{\left( b^{\prime }-4\right) r+3b}{2r^{2}}%
F^{\prime }-\left( 1-\frac{b}{r}\right) F^{\prime \prime },
\end{equation}%
\begin{equation}
p=-\frac{f}{2}+\frac{2r^{2}\left( b-r\right) \left( \Phi ^{\prime 2}+\Phi
^{\prime \prime }\right) +\left( r\Phi ^{\prime }+2\right) \left( b^{\prime
}r-b\right) }{2r^{3}}F+\frac{\left( r-b\right) \left( r\Phi ^{\prime
}+2\right) }{r^{2}}F^{\prime },
\end{equation}%
\begin{equation}
q=-\frac{f}{2}+\frac{2r\left( b-r\right) \Phi ^{\prime }+rb^{\prime }+b}{%
2r^{3}}F-\frac{\left[ -2r\left( r-b\right) \Phi ^{\prime }+\left( b^{\prime
}-2\right) r+b\right] }{2r^{2}}F^{\prime }+\left( 1-\frac{b}{r}\right)
F^{\prime \prime }.
\end{equation}%
The case of Ref. \cite{5} is easily observed if one sets $\Phi =$constant
and therefore%
\begin{equation}
\rho =\frac{f}{2}+\frac{\left( b^{\prime }-4\right) r+3b}{2r^{2}}F^{\prime
}-\left( 1-\frac{b}{r}\right) F^{\prime \prime },
\end{equation}%
\begin{equation}
p=-\frac{f}{2}+\frac{\left( b^{\prime }r-b\right) }{r^{3}}F+\frac{2\left(
r-b\right) }{r^{2}}F^{\prime },
\end{equation}%
\begin{equation}
q=-\frac{f}{2}+\frac{rb^{\prime }+b}{2r^{3}}F-\frac{\left( b^{\prime
}-2\right) r+b}{2r^{2}}F^{\prime }+\left( 1-\frac{b}{r}\right) F^{\prime
\prime }.
\end{equation}%
The other limit can be checked for $r=r_{0}$ ($b=b_{0}$) and the results
found in (33)-(35) are recovered.

Next, let's consider an isotropic velocity distribution which implies $p=q$
together with the ansaetze $b(r)=r_{0}^{2}/r$ and $\Phi \left( r\right)
=2\ln \left( \frac{r}{r_{0}}\right) .$ These in turn lead to%
\begin{equation}
r^{2}\left( r^{2}-r_{0}^{2}\right) F^{\prime \prime }+r\left(
2r_{0}^{2}-r^{2}\right) F^{\prime }+4r_{0}^{2}F=0.
\end{equation}%
This equation admits two independent solutions and to keep our calculation
analytic, we choose the simpler one given by%
\begin{equation}
F=C_{1}\sqrt{1-\frac{r_{0}^{2}}{r^{2}}},
\end{equation}%
in which $C_{1}$ is an integration constant. Same ansatz in the Ricci scalar
gives 
\begin{equation}
R=\frac{6\left( r_{0}^{2}-2r^{2}\right) }{r^{4}}
\end{equation}%
which implies (through $F=\frac{df}{dR}=\frac{f^{\prime }}{R^{\prime }}$)%
\begin{equation}
f=\frac{24C_{1}}{5r_{0}^{2}}\left( 1-\frac{r_{0}^{2}}{r^{2}}\right)
^{5/2}+C_{2}
\end{equation}%
for the integration constant $C_{2}$ that is related with the cosmological
constant. Finally one obtains 
\begin{equation}
f\left( R\right) =\frac{24C_{1}}{5r_{0}^{2}}\left( 1+\frac{\left( \frac{%
r_{0}^{2}R}{6}\right) }{1+\sqrt{1+\left( \frac{r_{0}^{2}R}{6}\right) }}%
\right) ^{5/2}+C_{2}.
\end{equation}%
We note that the Ricci scalar $R$ satisfies $\frac{-6}{r_{0}^{2}}<R<0$. This
causes no restriction on the above expression and is well defined everywhere
for $r\geq r_{0}.$ Our next step is to find the stress-energy components
which are as follow:%
\begin{eqnarray}
\rho &=&\frac{\left(
24r^{6}-12r^{4}r_{0}^{2}-18r^{2}r_{0}^{4}+6r_{0}^{6}\right)
C_{1}+5r^{5}r_{0}^{2}C_{2}\sqrt{r^{2}-r_{0}^{2}}}{10r_{0}^{2}r^{5}\sqrt{%
r^{2}-r_{0}^{2}}}, \\
p &=&q=-\frac{\left(
24r^{6}-52r^{4}r_{0}^{2}+32r^{2}r_{0}^{4}-4r_{0}^{6}\right)
C_{1}+5r^{5}r_{0}^{2}C_{2}\sqrt{r^{2}-r_{0}^{2}}}{10r_{0}^{2}r^{5}\sqrt{%
r^{2}-r_{0}^{2}}}.
\end{eqnarray}%
Furthermore the WEC implies that $\rho \geq 0$ and $\rho +p\geq 0,$ which in
closed form read%
\begin{equation}
\rho \geq 0\rightarrow \left(
24r^{6}-12r^{4}r_{0}^{2}-18r^{2}r_{0}^{4}+6r_{0}^{6}\right)
C_{1}+5r^{5}r_{0}^{2}C_{2}\sqrt{r^{2}-r_{0}^{2}}\geq 0
\end{equation}%
and 
\begin{equation}
\rho +p\geq 0\rightarrow \frac{C_{1}\left(
4r^{4}-5r^{2}r_{0}^{2}+r_{0}^{4}\right) }{r^{5}\sqrt{r^{2}-r_{0}^{2}}}\geq 0.
\end{equation}%
One can see that both conditions are satisfied if both $C_{1}$ and $C_{2}$
remain positive. In addition to WEC it is remarkable that the strong energy
condition (SEC) i.e. $\rho +p\geq 0$ and $\rho +3p\geq 0,$ is also satisfied
under the following condition:%
\begin{equation}
\frac{C_{2}}{C_{1}}\leq \frac{3}{5}\frac{3-16\xi ^{2}+8\xi ^{4}}{%
r_{0}^{2}\xi ^{5}}\sqrt{\xi ^{2}-1}.
\end{equation}%
Here $C_{1}$ and $C_{2}$ both are positive as chosen above for WEC and $\xi
\left( =\frac{r}{r_{0}}\right) $ is a positive parameter such that $1<\xi <%
\frac{\sqrt{4+\sqrt{10}}}{2}$. It should be added that even for the case of $%
C_{2}=0$, the SEC holds true in the particular range of $\xi .$ Let us add
that an expansion of $f(R)$ in powers of $R$ yields%
\begin{equation}
f\left( R\right) \cong \frac{24C_{1}}{5r_{0}^{2}}+C_{2}+C_{1}R+\mathcal{O}%
(R^{2})
\end{equation}%
which gives the exact combination of integration constants ($C_{1},C_{2}$)
that can be identified as the cosmological constant. The weak-curvature
expansion (58) suggests that $C_{1}\neq 0,$ is the crucial constant in order
to attain Einstein-Hilbert term, but $C_{2}$ can be disposed.

Finally we look at the wormhole's line element 
\begin{equation}
ds^{2}=-\left( \frac{r}{r_{0}}\right) ^{4}dt^{2}+\frac{dr^{2}}{1-\frac{%
r_{0}^{2}}{r^{2}}}+r^{2}\left( d\theta ^{2}+\sin ^{2}\theta d\varphi
^{2}\right)
\end{equation}%
and check the conditions which $b\left( r\right) =\frac{r_{0}^{2}}{r}$ and $%
\Phi \left( r\right) =2\ln \left( \frac{r}{r_{0}}\right) $ must satisfy. The
first condition which states that $\frac{b-b^{\prime }r}{b^{2}}>0$ is
satisfied since%
\begin{equation}
\frac{b-b^{\prime }r}{b^{2}}=\frac{2r}{r_{0}^{2}}>0.
\end{equation}%
The second condition implies that $b^{\prime }\left( r_{0}\right) <1.$ This
in turn reads $-r_{0}<1$ and is obvious. Finally $1-\frac{b}{r}>0$ means $1-%
\frac{r_{0}^{2}}{r^{2}}>0$ and is also satisfied for $r>r_{0}.$ The
condition on $\Phi \left( r\right) $ implies that it must be finite
everywhere without root for $r>r_{0}$ which is trivially satisfied (except
for $r\rightarrow \infty $). To complete our example we give the embedded
surface as $z\left( r\right) =r_{0}\cosh ^{-1}\left( \frac{r}{r_{0}}\right)
. $ This is shown in Fig. 1 for $r_{0}=1.$

\section{Conclusion}

One of the most challenging problems in wormhole physics is to find an
acceptable energy-momentum that will provide the outward push against
gravitational collapse. In Einstein's general relativity, i.e. $f(R)=R$,
this has not been possible. Now, with the advent of modified theories,
namely the $f(R)$ gravity, this may turn into reality. Our approach to the
problem of wormholes is to employ necessary existence conditions analogous
to black holes. The existence problem of horizon in a black hole plays the
similar role for throat radius in a wormhole. Expansion of metric functions
around the throat and substitutions into Einstein's equations derive the
necessary conditions. From these conditions, at the lowest order, we
determine the metric functions that yield a traversable wormhole. In this
process we have assumed that $f\left( R\right) $ admits non-zero derivatives
at least to third order, i.e. $\frac{d^{3}f}{dR^{3}}\neq 0$. With the
introduction of external matter sources into $f(R)$ we show that the weak
energy condition (WEC) (at least) is satisfied in the construction of a
wormhole in $f(R)$ gravity. The $f(R)$ function is explicitly determined
(Eq. (52)) in which the Ricci scalar satisfies $\frac{-6}{r_{0}^{2}}<R<0$.
Our results apply also to the case with / without a cosmological constant
since the latter arises automatically as a combination of integration
constants. In conclusion, in a generic class of $f(R)$ theories satisfying
the necessary conditions for existence of traversable wormhole, solutions
can be supported by physical stress-energy tensor. The embedding diagram for
the obtained wormhole is shown in Fig. 1.

\bigskip

\bigskip \textbf{Figure Caption:}

Fig. 1: The embedding surface (i.e. $dr^{2}+dz^{2}=\frac{dr^{2}}{1-\left( 
\frac{r_{0}}{r}\right) ^{2}}$ and $\theta =\frac{\pi }{2}$) for $b\left(
r\right) =\frac{r_{0}^{2}}{r}$ and $\pm z\left( r\right) =r_{0}\cosh
^{-1}\left( \frac{r}{r_{0}}\right) $ for $r_{0}=1.$


\begin{thebibliography}{99}
\bibitem{1} S. Nojiri and S. D. Odintsov, Phys. Rep. \textbf{505, }59 (2011);

A. De Felice, S. Tsujikawa, Living Rev. Rel. \textbf{13}, 3 (2010);

T. P. Sotiriou and V. Faraoni, Rev. Mod. Phys. \textbf{82}, 451 (2010);

L. Hollenstein and F. S. N. Lobo, Phys. Rev. D \textbf{78}, 124007 (2008);

S. H. Mazharimousavi and M. Halilsoy, Phys. Rev. D \textbf{84}, 064032
(2011);

S. H. Mazharimousavi, M. Halilsoy and T. Tahamtan, Eur. Phys. J. C. \textbf{%
72}, 1851 (2012).

\bibitem{2} A. DeBenedictis and D. Horvat, Gen Relativ Gravit, (2012) in
press;

M. A. Oliveira, arXiv:1107.2703v1.

K. A. Bronnikov, M. V. Skvortsov and A. A. Starobinsky, Grav. Cosmol. 
\textbf{16}, 216 (2010);

N. Furey, A. DeBenedictis, Class.Quant.Grav. \textbf{22}, 313 (2005);

F. S. N. Lobo, AIP Conf. Proc. \textbf{1458, }447\textbf{\ }(2011);

N. M. Garcia and F. S. N. Lobo, Phys. Rev. D \textbf{82}, 104018 (2010);

F. S. N. Lobo, Class. Quant. Grav. \textbf{25}, 175006 (2008).

\bibitem{3} S. E. P. Bergliaffa and Y. E. C. de O. Nunes, Phys. Rev. D 
\textbf{84}, 084006 (2011).

\bibitem{4} M. S. Morris, K. S. Thorne and Ulvi Yurtsever, Phys. Rev. Lett. 
\textbf{61, }1446\textbf{\ } (1988).

M. S. Morris and K. S. Thorne, Am. J. Phys. \textbf{56,} 395 (1988).

\bibitem{5} F. S. N. Lobo and M. A. Oliveira, Phys. Rev. D \textbf{80},
104012 (2009).
\end{thebibliography}
\end{document}